\documentclass[sigconf]{acmart}

\usepackage{comment}
\usepackage{csquotes}
\usepackage[htt]{hyphenat}

\def \system {PlanFitting}

\usepackage{multirow}
\usepackage{color}
\usepackage{colortbl}
\usepackage{stfloats}
\usepackage{longfbox} 
\usepackage{subcaption}
\usepackage{enumitem}
\usepackage{graphicx}
\usepackage{subcaption}
\usepackage{tabularx}
\usepackage{longtable}

\newcommand{\eg}{\textit{e.g.}}
\newcommand{\ie}{\textit{i.e.}}
\newcommand{\cf}{\textit{c.f.}}
\newcommand{\etal}{\textit{et al.}}

\newcommand{\labelphantom}[1]{%
  \parbox{0pt}{\phantomsubcaption\label{#1}}%
}

\definecolor{revisedcolor}{RGB}{0,0,255}

\newcommand{\hr}{\noindent\rule{\linewidth}{0.5pt}}

\newcommand{\ipstart}[1]{\vspace{2mm} \noindent{\textit{#1.}}}

\newcommand{\circledigit}[1]{\textbf{\normalsize{\textsf{\textcircled{\footnotesize{#1}}}}}}

\definecolor{tableheader}{HTML}{EFEFEF}
\definecolor{tablegrayline}{HTML}{d0d0d0}

\newcommand{\tabitem}{\textbullet~~}
\newcommand{\tablistitem}[2]{\tabitem \parbox[t]{#1}{#2}}

\newlist{itemtt}{itemize}{1}
\setlist[itemtt,1]{label={\textbullet},before={\ttfamily}}

\newcolumntype{L}[1]{>{\raggedright\let\newline\\\arraybackslash\hspace{0pt}}m{#1}}

\newcommand{\leftcell}[2][l]{%
  \begin{tabular}[#1]{@{}l@{}}#2\end{tabular}}

\AtBeginDocument{%
  \providecommand\BibTeX{{%
    \normalfont B\kern-0.5em{\scshape i\kern-0.25em b}\kern-0.8em\TeX}}}

\copyrightyear{2025}
\acmYear{2025}
\setcopyright{rightsretained}
\acmConference[CUI '25]{Proceedings of the 7th ACM Conference on Conversational User Interfaces}{July 8--10, 2025}{Waterloo, ON, Canada}
\acmBooktitle{Proceedings of the 7th ACM Conference on Conversational User Interfaces (CUI '25), July 8--10, 2025, Waterloo, ON, Canada}
\acmDOI{10.1145/3719160.3736607}
\acmISBN{979-8-4007-1527-3/2025/07}

\author{Donghoon Shin}
\authornote{Donghoon Shin conducted this work as a research intern at NAVER AI Lab.}
\orcid{0000-0001-9689-7841}
\email{dhoon@uw.edu}
\affiliation{%
  \institution{University of Washington}
  \city{Seattle}
  \state{WA}
  \country{USA}}

\author{Gary Hsieh}
\orcid{0000-0002-9460-2568}
\email{garyhs@uw.edu}
\affiliation{%
  \institution{University of Washington}
  \city{Seattle}
  \state{WA}
  \country{USA}}

\author{Young-Ho Kim}
\orcid{0000-0002-2681-2774}
\email{yghokim@younghokim.net}
\affiliation{%
  \institution{NAVER AI Lab}
  \city{Seongnam}
  \state{Gyeonggi}
  \country{Korea}
}

\begin{document}

\title[Personalized Exercise Planning with Large Language Model-driven Conversational Agent]{\system{}: Personalized Exercise Planning with Large Language Model-driven Conversational Agent}

\begin{abstract}
Creating personalized and actionable exercise plans often requires iteration with experts, which can be costly and inaccessible to many individuals. This work explores the capabilities of Large Language Models (LLMs) in addressing these challenges. We present \system{}, an LLM-driven conversational agent that assists users in creating and refining personalized weekly exercise plans. By engaging users in free-form conversations, \system{} helps elicit users' goals, availabilities, and potential obstacles, and enables individuals to generate personalized exercise plans aligned with established exercise guidelines. Our study---involving a user study, intrinsic evaluation, and expert evaluation---demonstrated \system{}'s ability to guide users to create tailored, actionable, and evidence-based plans. We discuss future design opportunities for LLM-driven conversational agents to create plans that better comply with exercise principles and accommodate personal constraints.
\end{abstract}

\maketitle

\section{Introduction}

Despite the benefits of regular exercise, people often struggle to meet recommended physical activity guidelines~\cite{vankim2013vigorous, durstine2013chronic, tucker2011physical}. To facilitate regular exercise, many digital tools follow the approaches of personal informatics~\cite{Li10StageBasedModel}, including activity tracking~\cite{apple_watch, fitbit, garmin}, visualization~\cite{Li10StageBasedModel, anderson2007shakra, consolvo2008flowers, lin2006fish, Kim2021DataAtHand, consolvo2006design}, and self-reflection~\cite{Choe2017VisualizedSelf, Kim2021DataAtHand, kocielnik2018reflection} on activity data. However, existing tools place less focus on \textit{exercise planning}, leaving users to create their own workout schedules, which can be challenging without domain expertise, especially when tailoring it to personal lifestyle constraints~\cite{de2011more, xu2022understanding}. As a result, people rely on professional planners (\eg{}, personal trainers, medical practitioners); yet involving experts also presents various setbacks, such as high costs, inaccessibility, and lack of customization due to broad client bases~\cite{mahaffey, pelletier2020implementation, castrillon2020people, melton2008current}. To enhance personalization in exercise planning, a handful of works have explored crowdsourced and peer-supported planning~\cite{agapie2018crowdsourcing, agapie2016plansourcing}, but these approaches still require human effort, time, and financial investment, while also depending on users to articulate clear preferences upfront---often challenging without expert input.

One potential solution to tackle these challenges is to use LLM-driven conversational agents (CAs) to tailor exercise plans to individuals. With the availability and scalability that CAs offer, we posit that CAs driven by LLMs can guide users to create and continuously refine plans tailored to their individual contexts. Specifically, recent advances suggest the potential of LLM-driven CAs in collecting information for social needs (\eg{}, healthcare~\cite{kocielnik2021can}) and synthesizing information in the knowledge task (\eg{},~\cite{LangChain,chen2023agentverse,ma2024crafting}), through iterative turn-taking with the user. Highlighting these potentials, in this work, we explore how LLM-driven CAs can be used to help individuals craft and revise personalized exercise plans.

To that end, we first conducted formative interviews exploring current practices in creating personalized exercise plans and the challenges faced in professional planning contexts. From the interviews with professional exercise planners ($N=5$) and lay individuals (\ie{} clients; $N=8$) who have experience in setting up personalized exercise plans with planners, we characterized key steps in crafting personalized exercise plans---goal-setting, collecting availabilities and anticipated obstacles, prescribing plans, and iteration, while grounding the plans on the core high-level guidance suggested by existing exercise guidelines (\eg{}, ACSM~\cite{acsm, garber2011quantity}). Additionally, we found that planners often face difficulties integrating exercise prescriptions into the irregular schedules of clients, with limited incorporation of client input during the iterative process of revising plans.

Based on these insights, we designed and developed \system{}, an LLM-driven CA that assists lay individuals in creating and refining their personalized exercise plans grounded on guidelines through interactive dialogue. With a dynamic prompting approach, \system{} leads users to engage them in conversations that gather essential information about their constraints identified in our formative study (\ie{}, exercise goals, availabilities, and potential obstacles to adherence). Using this information, the agent recommends exercises from the dataset through the retrieval-augmented generation and presents the plan in the form of implementation intention (\ie, \texttt{IF-THEN} rules)~\cite{gollwitzer1999implementation}---a concise, flexible scheduling framework grounded in behavioral psychology that links user intentions to specific events without rigid time scheduling~\cite{cohen2008cost}, while aligning the plan with established exercise guidelines~\cite{acsm}.
 
We conducted a user study ($N=18$) where the participants formulated a weekly plan and refined it with \system{}. Our results found that \system{} effectively helped participants articulate personalized constraints while adapting to their unique chatting styles. Also, participants found \system{} to be useful and usable, and highlighted the agent's role in guiding them towards creating personalized and actionable plans. Additionally, our intrinsic evaluation revealed that the generated plans reliably followed the established exercise guidelines, and expert planners ($N=3$) who evaluated the generated plans based on the exercise principle (\ie, FITT~\cite{desimone2019tortoise}) evaluated the \textit{frequency}, \textit{intensity}, and \textit{time} composition of the generated plans to be above average. However, they also identified opportunities to enhance the combination of exercise \textit{types}. Based on qualitative feedback from participants and expert planners, we also explore design implications for improving the use of LLM-driven CAs in creating personalized exercise plans.

\vspace{5pt}\noindent{}The main contributions of our work, along with the corresponding sections in the paper, are as follows:

\begin{itemize}[leftmargin=14pt, topsep=1pt]
    \item We present the results of our formative study, revealing the process and challenges of exercise planning between clients and expert planners, which informed the design of our conversational agent~(\S\ref{sec:formative_study});
    \item We introduce \system{}, an LLM-driven conversational agent that assists users in creating and refining personalized exercise plans. We present the agent's operationalization---including dialogue management and the interaction between the conversational agent and user---towards creating personalized and guideline-informed exercise plans~(\S\ref{sec:design});
    \item We present empirical findings from (i) a user study exploring how users interact with and perceive \system{}, (ii) an intrinsic evaluation assessing how well the generated plans follow established exercise guidelines, and (iii) an expert evaluation assessing the quality of the generated plans~(\S\ref{sec:user_study}, \S\ref{sec:user_study_results}).
\end{itemize}
\section{Related Work}
Our work builds on prior research that explored (1) personalized exercise planning, (2) technology-mediated exercise support and planning, and (3) LLM-driven conversational agents.

\subsection{Crafting Personalized and Actionable Exercise Plans}

Engaging in regular physical activity is essential for a healthy lifestyle; however, many people struggle to integrate sufficient exercise into their daily lives~\cite{de2011more, xu2022understanding}. To address this, establishing and adhering to exercise plans has proven effective for motivating individuals to sustain consistent physical activity~\cite{luszczynska2011self, gollwitzer1999implementation}. In response, several evidence-based guidelines have been proposed; for instance, the American College of Sports Medicine (ACSM)~\cite{acsm, garber2011quantity} and the U.S. Department of Health and Human Services~\cite{us_physical_activity, piercy2018physical} have created widely accepted guidelines that health professionals utilize to formulate effective exercise regimens. These guidelines provide general recommendations for planning (\eg{}, advising a minimum of 150 minutes of moderate-intensity exercise per week)~\cite{acsm, garber2011quantity, us_physical_activity, piercy2018physical}, along with definitions of exercise-related terms. With these comprehensive guidelines, individuals can further tailor their exercise plans to their preferences and constraints, which is known to contribute to the successful adoption of plans, though achieving such personalization is not trivial and often requires the expertise of exercise professionals.

Another line of research in behavioral psychology and sports medicine has explored the effective intervention and format of exercise prescriptions. One well-known approach is \textit{implementation intention}, which comprises a specific plan linking a particular circumstance to corresponding actions~\cite{gollwitzer1999implementation, Hagger2013ImplementationIntention}. Formatted as \texttt{IF-THEN} rules, implementation intentions are often combined with action planning by including environmental cues~\cite{Hagger2013ImplementationIntention}. For example, one can set up an exercise plan like ``\texttt{IF} \textit{I come back home in the evening}, \texttt{THEN} \textit{I will jog for 30 minutes.}'' By effectively transforming intentions into actionable steps, implementation intentions have demonstrated success in various behavior change contexts (\eg{}, managing a healthy diet~\cite{adriaanse2011implementation, achtziger2008implementation, gratton2007promoting, reuter2008dietary}, reducing bedtime procrastination~\cite{valshtein2020using}, smoking cessation~\cite{mcwilliams2019beyond, conner2010long}). Likewise, in the context of exercise, implementation intentions have been shown to be effective in promoting physical activities~\cite{lippke2004initiation}, suggesting its adaptability to personalized exercise planning.

\subsection{Technology-mediated Exercise Support and Planning}

Given the importance and barriers of regular exercise, researchers in HCI have long investigated the design of digital tools to facilitate the tracking of physical activity~\cite{klasnja2011evaluate, consolvo2006design, lin2006fish, kocielnik2018reflection, anderson2007shakra, consolvo2008flowers, kim2017omnitrack, kim2022mymove, Kim2021DataAtHand}. These tools commonly incorporated a personal informatics and self-tracking approach, where the tool provides insights about the user's progress and status of exercise so that they stay motivated and knowledgeable about themselves~\cite{Li10StageBasedModel}. For example, UbiFit Garden employed metaphoric visualization of various daily exercise metrics (\ie, exercise categories and amount) to help users keep up with the progress of activity at a glance~\cite{consolvo2008flowers}. Reflection Companion engages users in a daily SMS dialogue that promotes self-reflection on their physical activity levels captured by activity trackers~\cite{kocielnik2018reflection}.

Meanwhile, research on technology support for exercise planning is relatively sparse, with only a few works exploring planning tools (\eg, \cite{lee2015personalization, xu2022understanding, agapie2016plansourcing, agapie2018crowdsourcing}). Xu \etal{} investigated digital planning experiences for physical activities~\cite{xu2022understanding}, and Lee \etal{} probed the effect of reflective strategies in physical activity planning~\cite{lee2015personalization}. Agapie~\etal{} proposed involving peers and crowdworkers to help generate custom exercise plans for health behavior change~\cite{agapie2018crowdsourcing, agapie2016plansourcing}. These works showed technology's potential in constructing personalized exercise plans, yet they still require substantial human involvement for plan formulation which limits the sustainability and scalability. Although a handful of commercial applications (\eg{},~\cite{fitbod, fitnessai}) have attempted to use AI for planning resistance training, they are limited to maximizing the effectiveness of muscle growth, and lack adaptability to individual schedules. To address these limitations, our work proposes leveraging the scalability of conversational agents by exploring their use in supporting iterative planning to assist users in creating personalized exercise plans.

\subsection{Leveraging LLM-driven Conversation Agents for Personalized Exercise Planning}

Conversational agents (CAs) have found widespread use in gathering information for various social purposes, such as web surveys~\cite{kim2019comparing} and fostering self-disclosure~\cite{lee2020hear, park2021designing}, as they are easily scalable and readily available to use. Their applications extend to healthcare and well-being contexts, where prior works in healthcare leveraged CAs to collect health-related information from users~\cite{erazo2020chatbot, ben2022assessing}, and have been shown to be a preferred way of providing social needs information---especially for those with lower health literacy~\cite{kocielnik2020harborbot, kocielnik2021can}. However, conventional rule-based CAs are typically known to suffer from constrained interaction capabilities, lack of extensibility to other domains once designed, and rigid input demands, particularly because they lack robust natural language adaptability and comprehension~\cite{wei2023leveraging, abd2021perceptions, jain2018evaluating}. Moreover, analyzing the information gathered by these agents to compose meaningful insights still relies heavily on human practitioners, resulting in substantial manual efforts.

Recent advancements in LLMs demonstrate the potential of CAs to engage more fluidly in dialogue while synthesizing information to deliver valuable insights in real-time. By incorporating detailed behavioral guidelines (\ie{}, preprompt/system prompt), these agents can adapt their conversational style to suit diverse contextual requirements, enabling open-ended interactions without the need for extensive training dialogue corpora. This mechanism has streamlined bootstrapping in novel conversational topics, as evidenced by the broad range of applications from general-purpose agents (\eg{}, ChatGPT~\cite{chatgpt}, Gemini~\cite{gemini}) to specialized research prototypes (\eg{}, health data collection~\cite{wei2023leveraging}, recommender system~\cite{friedman2023leveraging, chen2023palr}). Moreover, innovations in LLM frameworks (\eg{}, LangChain~\cite{LangChain}, AgentVerse~\cite{chen2023agentverse}) and applications (\eg{},~\cite{ma2024crafting}) have further demonstrated how LLM-driven agents can be harnessed to perform complex, knowledge synthesis tasks (\eg{}, data-driven question-answering). In the context of exercise planning, this suggests the possibility of harnessing LLM-driven CAs to not only flexibly collect individual users' constraints but also to integrate and analyze these to create cohesive plans.

Despite these, ensuring an LLM has learned specific knowledge during its pretraining remains challenging~\cite{Kandpal2023LLMLongTailKnowledge}. This has been pointed out to make LLM-based CAs prone to returning errors, particularly regarding domain-specific conversations that require specialized knowledge~\cite{gao2024rag, Kandpal2023LLMLongTailKnowledge, Singhal2023LLMEncodeClinical}--including exercise planning. For example, when tasked with generating exercise plans, the CA may offer recommendations for exercise types and amounts that lack evidence, especially with regard to an individual's unique situation. One potential approach to enhance the accuracy and credibility of such conversations is Retrieval-Augmented Generation (RAG), in which critical knowledge required for the task is retrieved from an external knowledge base and incorporated into the preprompt to augment the LLM agent's response generation~\cite{gao2024rag, shuster2021rag}.

Drawing inspiration from research on using CAs for social needs and the adaptability of LLMs, this work investigates how LLM-driven CAs can be designed to interact with users towards the creation of personalized plans. More specifically, we aim to enable the free-form expression of user constraints and requirements for exercise planning, and synthesize them into evidence-based plans. To enhance the robustness of following up the dialogue context, we also propose a design choice to implement a separate LLM routine for the agent that generates a dialogue summary, which is injected into the preprompt of the LLM for conversation. Lastly, we incorporate the retrieval mechanism by making the agent refer to an external exercise database to avoid recommending seemingly plausible but irrelevant (\ie, ``hallucinated''~\cite{gao2024rag}) exercise types. In our work, we demonstrate how these mechanisms can holistically support personalized exercise planning that is reliable and grounded on credible knowledge.

\section{Formative Study}\label{sec:formative_study}

To understand the current practice of conducting personalized exercise planning and the challenges that arise during the process, we conducted a formative interview study with exercise planners ($N=5$) and clients ($N=8$). The study protocol was reviewed and approved by the institutional review board.

\ipstart{Exercise planners}
From an in-house clinic and a corporate internal network, we recruited five experts (FP1--5; three females and two males) who are experienced in setting up personalized exercise plans for clients. Of all, three were physical therapists, another was a physiatrist, and the other was a kinesiologist. On average, they had 9.8 years ($SD=4.5$) of experience in advising and planning exercise planning.

\ipstart{Clients}
We recruited eight individuals (FC1--8; 6 females and 2 males) by advertising our study on a local community platform and the corporation's internal bulletin boards. We required participants to have experience setting up their personalized exercise plans under the advice of exercise experts (\eg{}, clinicians, physical therapists, personal trainers, etc.). Clients were aged between 26 and 45 ($M=35.0$); three participants responded that they have/had engaged in exercise under the personalized exercise plans for less than three months, three participants for 3 to 6 months, and the other two participants for more than six months.

We invited each participant to a 1-hour semi-structured interview session. During the session, we asked each exercise planner to primarily share insights into (1) their planning procedures for clients and (2) the challenges they encountered while setting up personalized plans for/with the clients. Likewise, clients were prompted to elaborate on (1) their experiences and process of planning exercises with exercise planners and (2) the challenges they faced during the planning. Each interview was audio-recorded and later transcribed, and we compensated 50,000 KRW (approximately 35 USD) and 30,000 KRW (approximately 21 USD) for each planner and client, respectively.

After the interview, we analyzed the interview transcripts using thematic analysis~\cite{braun2006using}. The analysis was done in a bottom-up approach, where the two authors first familiarized themselves with the raw responses independently. Then, each author identified emerging themes from the responses, brought these themes to a regular meeting, and compared the themes until they reached a consensus. As a result, we could derive the final themes as detailed in \autoref{formative:practice}~and~\ref{formative:challenges}.

\subsection{Practice of Personalized Exercise Planning} \label{formative:practice}

First, planners reported that they primarily inform the exercise plans with globally recognized guidelines (\eg{}, ACSM guidebook~\cite{acsm}), which emphasizes engaging in a minimum of 150 minutes of moderate-intensity exercise per week. However, they suggested that these guidelines do not provide specific guidance on tailoring to individuals' varying lifestyles: \textit{``Actually, even if you take a look at those exercise planning guidebooks, there won't be anything more detailed than [showing a page that defined some case studies of individuals] (...) that's the end of ‘evidence-based’ personalization.''}~(FP4) As such, planners use them as a flexible framework rather than strict rules, making tailored modifications while adhering to such high-level principles: \textit{“I'm just following a broad guide and customizing a lot in that scope. Shouldn't the details within it be personalized?”}~(FP2)

More specifically, we could characterize the process of personalization, and surface the common information that planners gather from clients in this process to tailor plans to their lifestyles, such as personal goals for the exercise, personal obstacles, and feedback (during the follow-up sessions), delivered through either verbal communication or a combination of a survey form and oral report:

\ipstart{Understanding client's main goals for exercise}
Every planner responded that they begin by identifying the client's goal for the exercise, highlighting the importance of defining the purpose and setting clear objectives to motivate clients. To enable this, they engaged in conversations with clients to find out their own necessity and benefits of exercise to enhance motivation, particularly for newcomers: \textit{“For managing exercise plans, it's crucial to first motivate by discussing goals first rather than just telling them to do it.”}~(FP1)

\ipstart{Surfacing available amount of times for exercise and potential obstacles}
Once identifying the goals for the exercise, planners are reported to ask clients questions about their availabilities, such as how much time they would be available to spare for exercise: \textit{“For those who don't have set regular office hours or for nurses working 3/4 shifts, I ask and look at how much personal time the client can exercise on a regular basis.”}~(FP3) Also, planners surface factors from clients that may potentially make it challenging for them to exercise during those times (\eg{}, physical constraints, parenting), to make the exercise planning more viable and realistic: \textit{“I told my planner when my menstrual cycle comes (...) And (as a developer in a company) I told them whenever there is a schedule for releasing a new version that my condition won't be good for about three following days.”}~(FC5)

\ipstart{Prescribing plans}
Based on collected exercise goals, availabilities, and obstacles, planners create a personalized exercise plan for clients. While planners are willing to provide detailed plans down to specific times, the limited availability of planners makes this approach impractical: \textit{“I can't do detailed time planning (...) It seems inconsistent (with my current availability) to generate highly detailed plans, like scheduling at a certain time.”}~(FP4) As a result, planners and clients typically receive a weekly exercise plan with recommended days and hours, exercise types, allowing clients to exercise at their own convenience to meet their requirements: \textit{“They (planner) didn't ask me to exercise at a specific time; they just told me to do a certain amount of some exercises during the week.”}~(FC4)

\ipstart{Revisit regularly (\eg{}, weekly, bi-weekly) to share feedback and iterate on the plan}
Emphasizing the importance of viewing the exercise planning as a feedback-driven iteration, rather than a one-time interaction, planners and clients revisit the plans regularly (\eg{}, weekly, bi-weekly) to check if the exercises need to be modified: \textit{“There are types of exercise that go in and out (...) After solving the urgent problem, if I wanna get a nicer body shape, other exercises may go in or out.”}~(FC1) Gathering newly emerged feedback and constraints, planners make adjustments to exercise types and/or duration: \textit{“Clients first give it a try, and I gather feedback when they come back in the following week based on their experience trying the exercise plan. If they think it won't work for any reason, I ask them to let me know, and we can start the revisions from there, just like forming and iterating on a hypothesis.”}~(FP2)

\subsection{Challenges of Personalized Exercise Planning}\label{formative:challenges}

\subsubsection{Difficulty of contextualizing the exercise within their own schedule}

After the prescription of a broadly defined weekly exercise plan, clients are required to incorporate these exercises into their own schedules by themselves. However, clients from our interviews reported that such an ‘autonomous’ process, without clearer support on identifying when to exercise within their actual schedule, makes it difficult for them to cope with unexpected variables (\eg{}, appointments, work schedules). As such, adhering to the plans becomes highly reliant on their own motivation, making clients prone to becoming complacent: \textit{“I think it's mostly about getting the number of exercises and then performing them on my own, so my own willingness is the most important factor (...) If I suddenly have to work at night, I just end up not doing exercise that day because there's no one pushing me to do and I feel like I can just do it later.”}~(FC2)

In particular, these issues are reported to worsen over time. As time passes, various triggers that may lower motivation are reported to emerge, such as moments of stagnation during their exercise progress, which is exacerbated over time, leading to a tendency to continuously postpone or skip prescribed exercise: \textit{“If you aim for a weight loss, there are times when you reach a point where you're not losing any more weight (...) then my motivation decreased a bit, so sometimes I took a day or two off, rested a bit more, or skipped it in various other ways. So, I'm skipping more than I did in the beginning.”}~(FC6)

\subsubsection{Limited availability of planners affecting the iteration process and adaptation to fluctuating schedules}

Clients expressed struggles around accommodating sudden, unexpected time changes caused by their irregular lifestyles and work schedules. In these cases, reaching out to planners for real-time schedule adjustments is unrealistic, due to the other personal/work commitments planners have. As a result, clients reported their desire for more flexibility in communication when iterating on exercise plans: \textit{“I have been meeting my planner every week (...) it was sad to see whenever I have a schedule change and need an alternative, I couldn't ask about the plan iterations right away for the other days.”}~(FC6)

Such an issue, ironically, is reported to make the whole exercise schedule of clients even more dependent on the planners' decision-making process. Consequently, if the weekly meeting is canceled as either the client or the planner is unable to attend the weekly meeting, it often results in a disruption of the exercise for the entire week: \textit{“There were instances when the trainers were not available due to their other commitments (...) the whole exercise for the following week messed up.”}~(FC5)

\subsubsection{Limited adaptability of planners in engaging with and incorporating client feedback}

Even when meeting to discuss the exercise plan, clients often struggle to have their concerns and input incorporated into the plans. Indeed, clients shared several anecdotes when they felt their opinions were dismissed, or they had to spend a considerable amount of time advocating for their points to be considered: \textit{“You know, I can't see the planner every day and have to meet them face to face, and my daily conditions are different every day (...) but I always had to follow the same fixed program. I once went on a trip to [an attraction], but even when I explained this situation in advance my planner just asked me to keep exercising while traveling. It's too inflexible and feels too coercive.”}~(FC2)

The prescribed plan's inability to cater to unique constraints such as travel schedules could discourage clients from following through. In the worst case, disagreements stemming from this lack of flexibility have sometimes even led clients to discontinue their programs entirely: \textit{“I and planners had disagreements on the types of exercise, and I discontinued planning for the exercise with my personal trainer from that moment.”}~(FC5)
\section{\system{}}\label{sec:design}

Our formative study revealed the overall planning process of personalized exercise plans, as well as the challenges that emerge during the process. Building on these insights, we designed and implemented \system{}, a conversational agent system aimed to help individuals set up their personalized exercise plan and iterate on it. Focusing on the expressivity and comprehensibility that LLMs offer, we designed our system using LLMs to foster engaging interaction, while adapting to the unique constraints of users and allowing them to iterate their plans.

Informed by the planning procedure we surfaced from our preliminary study, we formulated the interaction process of \system{} into the following three stages: First, (1) the user provides exercise-related constraints (\ie{}, goals, availabilities, obstacles) to the agent. Then, the agent (2) offers a personalized exercise recommendation based on the provided constraints and (3) generates a personalized weekly plan. Lastly, (4) the user may revisit \system{} where the agent assists in refining the plan by accommodating the user's changing constraints. This way, we aimed to accommodate \system{} to general user groups by taking into account their constraints around performing exercises, comparing them with the list of exercises and the strength/relevant muscles involved, and having LLM associate them and recommend feasible exercises.
In the following, we describe the design of \system{}'s dialogue system with the underlying LLM pipeline, as well as its implementation.

\subsection{Interaction and Conversational UI Design}\label{sec:chat_flow}

\begin{figure*}
    \centering
    \includegraphics[width=\textwidth]{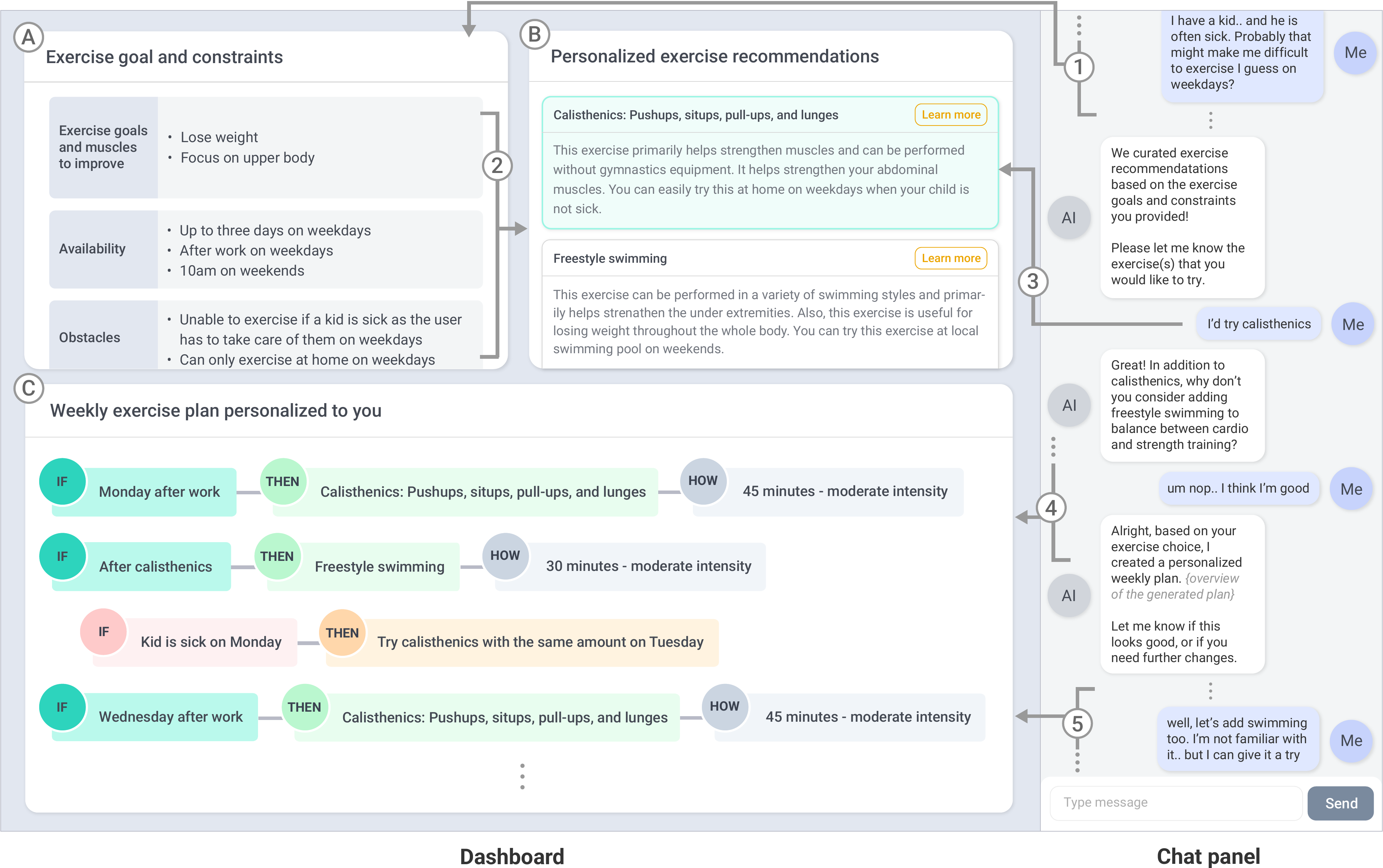}
    \caption{Key screen and interaction flow of \system{}. \circledigit{1} Once the user describes the goal of the exercise and their own constraints in a natural language on the chat panel, they are parsed and synchronized with the dashboard. \circledigit{2} Based on the collected information, \system{} recommends exercises and \circledigit{3} the user can provide the exercise type(s) they want to include. Once the user finalizes exercise types, \circledigit{4} the agent returns a weekly exercise plan, where the user can \circledigit{5} continuously iterate on the plan through natural language.}
    \Description{This figure illustrates the key screen and conversation/interaction flow of PlanFitting. Once the user describes the goal of the exercise and their own constraints in natural language on the chat panel to the conversational agent, this information is parsed and synchronized with the dashboard. Based on the provided information, the agent recommends exercises, allowing the user to provide the exercise type(s) they want to include. After the user finalizes exercise types, the agent returns a weekly exercise plan, which the user may further iterate on by interacting with the system.}
    \label{fig:interaction}
\end{figure*}

\system{} is designed as a web-based conversational UI (\autoref{fig:interaction}), where the users can primarily interact with the CA on the chat panel (\autoref{fig:interaction}-Chat panel) via natural language. It is supported by the dashboard providing an overview of the current status of the conversation (\autoref{fig:interaction}-Dashboard), summarizing the exercise goal and constraints (\autoref{fig:interaction}-\circledigit{A}), recommended exercise list from the system (\autoref{fig:interaction}-\circledigit{B}), and the exercise plans (\autoref{fig:interaction}-\circledigit{C}). Every information on the dashboard is automatically updated on every conversational turn so that the user can stay on track. In the following, we describe the detailed interaction flow between the CA and the user.

\subsubsection{Collecting exercise-related user constraints}

The CA first takes the lead by collecting essential information required for crafting a personalized exercise plan. Specifically, the CA proactively asks questions aimed at gathering the personal constraints of the user as follows:

\begin{enumerate}
    \item Exercise goals: The user's goal of exercise, either in a format of intended purpose or the specific muscle group they aim to target
    \item Availability: The user's available times for the exercise, either in the exact time format (\eg, `\textit{7 pm}') or in a descriptive form (\eg, `\textit{after work}')
    \item Potential obstacles: Any expected obstacles they anticipate that could potentially impede their exercise routine (\eg, `\textit{chance of working until late night}')
\end{enumerate}

\subsubsection{Exercise type recommendation}

After the user has shared all the necessary constraints, the agent proceeds to offer personalized exercise recommendations, where the system provides up to five exercise options based on the curated list of exercises from the predefined list of exercises. More specifically, we used the list of exercises from Agapie \etal{}~\cite{agapie2018crowdsourcing} that contains 112 common exercises that were curated by the expert exercise planners. The list contains the name of the exercise, as well as its alternative names (if any), intensity, laypeople description (\eg{}, definition, how to perform), and the muscles involved/exercise type. As the list is stored and loaded in CSV format, it can be easily expanded by altering with external exercise databases in the future if needed.

The recommended exercises are displayed on the dashboard with a brief description, which summarizes the definition of the exercise and the reasoning behind the recommendation (\autoref{fig:interaction}-\circledigit{B}). For users seeking more comprehensive information about a particular exercise, a ‘more’ button is provided where the users may click to retrieve additional details of the exercise from the attached database. Then, users are asked to select their desired exercises by either typing the name of the exercise(s) into the chat panel in a free form or clicking on them on the dashboard; if they wish to explore additional exercise options, they are also allowed to simply ask a request to the CA, which will result in refreshing the recommendations.

\subsubsection{Generating a personalized exercise plan}\label{sec:exercise_guideline}
After the user finalizes the exercise types, \system{} generates and outlines an exercise plan, with its structured format displayed on the dashboard (\autoref{fig:interaction}-\circledigit{C}).

\ipstart{Format of the plan}
Our interview study suggests that prescribing exercise broadly (\eg{}, specifying a weekly amount) could burden users with scheduling and possibly lower motivation. Thus, to better contextualize the exercise plan within the user's availabilities, \system{} offers each exercise plan in an \textit{implementation intention}~\cite{gollwitzer2006implementation} format, a grounded strategy rooted in behavioral psychology that aligns the user's intentions with specific events, hence offering a structured format in well-established \texttt{IF-THEN} statements. (\ie{}, \textit{“\texttt{IF} \{availability (time or situation)\}, \texttt{THEN} do \{exercise type\} for \{amount\} at \{intensity\}”})
In addition, the agent offers a \textit{coping plan} for each plan, which equips users with an alternative plan to follow when the original plan cannot be executed due to the obstacles that may happen. (\ie{}, \textit{“\texttt{IF} \{obstacle\}, \texttt{THEN} \{alternative\}”})

\ipstart{Grounding a plan to global exercise guidelines}\label{sec:global_guidelines}
To earn rigor for the generated plans, the agent applies a common set of guidelines that we elicited from the recommendations offered by the universally recognized exercises guidelines (\ie{}, ACSM~\cite{acsm, garber2011quantity}, U.S. Department of Health and Human Services~\cite{us_physical_activity, piercy2018physical}), as well as their previous application to the technology-mediated exercise planning~\cite{agapie2018crowdsourcing}:

First, the agent is instructed to allocate exercises totaling more than 150 minutes per week~\cite{acsm, garber2011quantity, us_physical_activity, piercy2018physical, agapie2018crowdsourcing}. To comply with the guidelines, it also accounts for vigorous-intensity exercises by doubling their allocated time when calculating the total exercise duration~\cite{us_physical_activity, piercy2018physical, agapie2018crowdsourcing}. In addition, to balance between cardio and strength training~\cite{acsm, garber2011quantity, agapie2018crowdsourcing}, if the user had initially chosen exercises of either type only, the agent asks users to consider incorporating both types of exercise. Lastly, the agent puts a minimum of a one-day rest period between exercise sessions, if the user constraints allow, to prevent any potential negative effects of consecutive days of exercising the same or adjacent muscle group~\cite{acsm, garber2011quantity, agapie2018crowdsourcing}.

\subsubsection{Revisiting and refining the exercise plan}

Following the initial planning phase, \system{} is designed to allow for iteration of the plans by inquiring users about their satisfaction with the existing plan, when the user returns to the system. More specifically, \system{} is instructed to first ask the user whether they followed the previous week's exercise plan and whether they were satisfied with it. If the user is satisfied with their plan, the agent asks if they are willing to extend the allotted time to adhere to the progression principle (\ie{}, gradually increase the engagement in exercise) of the exercise~\cite{acsm}. Otherwise, if the user indicates dissatisfaction, it solicits feedback on the specific aspects that require revision, facilitating an iterative approach to refining the plan. As such, the agent enables ongoing, open-ended planning, conducive to continuous improvement based on user input.

\vspace{5pt}\noindent{}In summary, the interaction flow between the user and the CA is structured to facilitate user engagement, provide exercise recommendations, and enable the creation of personalized exercise plans that adhere to recognized exercise guidelines, while allowing for the iteration of generated plans.

\subsection{Conversational Pipeline Design}

\begin{figure*}
    \centering
    \includegraphics[width=\textwidth]{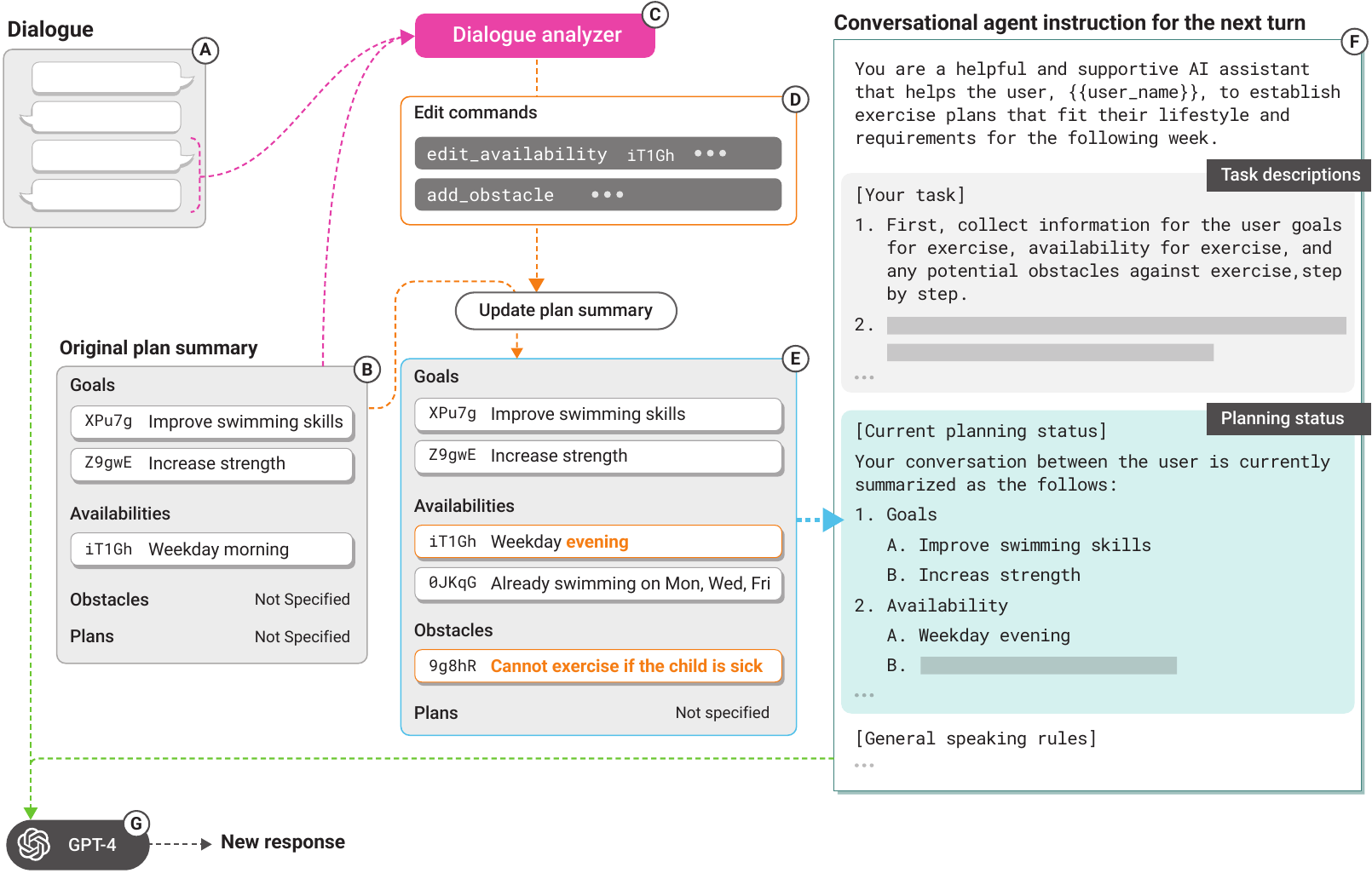}
    \caption{Illustration of how the \system{} computes and returns the next dialogue of the conversational agent and updates the dashboard based on the current dialogues}
    \Description{This figure illustrates how PlanFitting's conversational agent computes and returns the next dialogue and updates the dashboard based on the current dialogues.}
    \label{fig:prompting}
\end{figure*}

\autoref{fig:prompting} illustrates the pipeline of \system{}'s CA system.
\system{}'s CA is driven by two LLM components: a \textbf{response generator} (\autoref{fig:prompting}-\circledigit{G}) and the \textbf{dialogue analyzer} (\autoref{fig:prompting}-\circledigit{C}). The response generator generates the agent's response based on a global instruction (\autoref{fig:prompting}-\circledigit{F}) and the current dialogue (\autoref{fig:prompting}-\circledigit{A}). The user's constraints and generated plans are maintained in a data structure called ``plan summary'' (\autoref{fig:prompting}-\circledigit{B}), which maintains the current status while providing information to be displayed on the dashboard UI.

\ipstart{Plan summary update} Inspired by memory management techniques from the NLP discipline (\eg, \cite{bae2022keepmeupdated}), we designed the dialogue analyzer to generate edit commands that modify the previous state of the plan summary. The dialogue analyzer receives the latest turn pair (\ie, the CA's message and the user's response; \autoref{fig:prompting}-\circledigit{A}) and the plan summary of the previous cycle (\autoref{fig:prompting}-\circledigit{B}) as inputs and generates a list of edit commands (\eg, add, update, and remove; \autoref{fig:prompting}-\circledigit{D}) that reflect the changes caused by the new messages. Then, the system applies the edit commands to the plan summary and generates a new plan summary (\autoref{fig:prompting}-\circledigit{E}). The CA updates the plan summary every time before it generates and returns a response to the user. The base prompt of the dialogue analyzer is in Appendix~\ref{apdx:dialogue_analyzer}.

\ipstart{Conversation} Once the plan summary is updated, an instruction prompt (\autoref{fig:prompting}-\circledigit{F}) is formulated and fed into the response generator. The instruction includes the task descriptions on how to carry on the conversation (\autoref{fig:prompting}-\circledigit{F}, Task descriptions), and the current plan summary to inform the model with which constraints are missing, thus what needs to be asked in the following dialogues (\autoref{fig:prompting}-\circledigit{F}, Planning status).

When defining tasks for exercise type recommendation and generating plans, we established rules to append XML data to the message so that the system can systematically parse responses and integrate them into the user interface. For example, we specified the message rules for creating the plan as follows:

\hr{}
\begin{displayquote}
\texttt{Using the exercise types that the user selected, plan for and return the user's exercise plan in the implementation intention format\\...\\Each implementation intention rule should be accompanied by corresponding coping plans that can be plan B when the user fails to adhere to meet the main rules. It should assume the failure of each of the user's availabilities due to the obstacles the user mentioned\\...\\Each exercise/coping plan should be described in an IF-THEN format along with AMOUNT inside\\...\\(Example:\\<If>Monday after work</If>\\<Then>\\\hspace*{3mm}<Exercise>Running</Exercise> \\\hspace*{3mm}<Amount>60 minutes - moderate intensity</Amount>\\</Then> \\<If>After running</If> \\<Then>\\\hspace*{3mm}<Exercise>Pilates</Exercise>\\\hspace*{3mm}<Amount>30 minutes - vigorous intensity</Amount>\\</Then>\\<If>Too sleepy after work on Monday</If> <Then>\\\hspace*{3mm}<CopingPlan>Do the same exercises on Tuesday</CopingPlan>\\</Then>)}
\end{displayquote}
\hr{}


\vspace{5pt}\noindent{}To implement exercise recommendations, we employed function calling~\cite{function_calling} to extract exercise-related keywords from the user dialogue and enable structured data retrieval. Coupled with function calls, we used cosine similarity to compute the semantic closeness between the user's input and each predefined exercise description. We believed leveraging cosine similarity to be particularly suitable for this task, as it captures semantic similarity between vectors while making it robust to variation in user phrasing. To support this, we generated vector representations (\ie{}, word embeddings) using a pre-trained sentence embedding model (\texttt{text-embedding-ada-002}), which encodes both user input and exercise metadata (\ie{}, name, description, involved muscles) into a shared semantic space. By computing cosine similarity between these embeddings, the system identifies exercises that are semantically aligned with the user's intentions, even when those are expressed in informal or diverse language.

More specifically, we first embedded the title and description of each exercise from our prepared list and saved them as embedding vectors. Once the user finishes providing their constraints and an LLM detects if the exercise recommendation is needed, a function calling that takes the goal and obstacles as input and returns the recommended exercises is triggered. Here, the function is programmed to embed the detected exercise-related keywords of users as an embedding, which is then compared to each embedding of each exercise from the list to calculate the cosine similarity and return the types of exercise that have the top 5 cosine similarities to the user in a JSON format. Then, similar to how the system does for generating the exercise plan, \system{} formats the output JSON to the XML format through Regex postprocessing, which is then populated in the dashboard (\autoref{fig:prompting}-\circledigit{B}).

\subsection{Implementation}

\system{} system consists of two components: (1) a conversational UI and (2) a backend server, where the user interacts with the interface, whose chat is computed to return the response from the backend server.

The conversational UI is built as a web-based application on a JavaScript-based framework (SvelteKit). For the backend, we employed a Python Flask API server that takes the user's name and chat message as inputs and generates the subsequent message along with detected metadata--including exercise goals, availability, constraints, and recommended/selected exercise types. To run LLM computations in our conversational pipeline components, the system uses OpenAI~\cite{openaiAPI} GPT-4 model with the following parameters: $temperature= 0.5$, $top\_p=1$, $frequency\_penalty=0$, and $presence\_penalty=0$.
\section{User Study}\label{sec:user_study}

To gain a comprehensive understanding of the use of \system{}, we conducted a user study where participants interacted with \system{} to set up their exercise plans with the CA based on their own goals and constraints. In addition, the adherence to the guidelines and rigor of these plans was later assessed through intrinsic evaluation and expert evaluation, respectively. The study protocol was approved by our institutional review board.

\subsection{Participants}

We posted our study recruitment to a local online community platform and the corporate bulletin board, where we required participants to be (1) aged over 19, (2) motivated to do regular exercise, and (3) not currently doing exercise under the specific plan advised by planners to avoid any conflict, (4) who can participate in an in-person lab study. As a result, we recruited 18 participants (P1--P18; 11 females and 7 males) who were aged between 19 and 54 ($M=33.2$). Of all, six were full/part-time employees by the time they were participating in our study, six were college students, one was a retiree, and five responded that they were either stay-at-home parents or unemployed. We compensated 50,000 KRW (approximately 35 USD) as a gift card for their participation.

\subsection{Study Procedure \& Tasks}

To explore how the participants create and refine their exercise plans with \system{}, we structured the user study in the following phases: (1) initial planning, (2) iteration, and (3) debriefing. Throughout the planning, each participant was asked to think aloud in order for us to better surface their lively experience interacting with the CA.

\ipstart{Initial planning}
The initial phase involved participants being guided through the process of configuring their exercise plans with the assistance of \system{}. Participants were asked to interact with the agent to articulate and input their specific exercise goals, availabilities, and any potential obstacles. At the same time, they were also encouraged to freely ask questions to the agent and iterate on their plans until they were satisfied. As such, we aimed to mirror the process of tailoring exercise plans to individual constraints based on the overall guidance of the \system{} system.

\ipstart{Iteration}
After setting up their weekly exercise plan initially, participants were instructed to move on to the second phase. In this phase, they were asked to imagine themselves in the upcoming week, having completed their exercises successfully, and to also consider scenarios that may have hindered their progress in the previous weeks. To assist them in this process, we presented example scenarios for their reference (\eg, \textit{“I intended to swim last week, but I'd rather avoid such location-dependent activities due to the hassle of making reservations”}). In cases where they had nothing to change, they could engage with the system as if they were satisfied with their plan.

Once they had formulated their scenarios, participants were encouraged to use \system{} to review and fine-tune their exercise plans over a designated time frame. They were asked to freely describe adjustments to the agent that they would want to make, such as exercise availabilities, types, and amounts.

\ipstart{Debriefing}
During the final debriefing phase, we conducted a survey and a semi-structured interview with each participant to gather their feedback, insights, and reflections on both the planning process and their interactions with \system{}.

The survey was designed to assess their subjective evaluation of how personalized and actionable the plan they created with \system{} is, as well as their degree of acceptance and adoption of the \system{} system. To evaluate the level of personalization and actionability, we measured \textit{follow} and \textit{fit} for personalization, and \textit{specificity}, \textit{encouragement}, \textit{vocabulary}, and \textit{accuracy} for actionability on a 7-point Likert scale, following the rubric from the prior literature~\cite{agapie2018crowdsourcing} that evaluated the quality of the plan. For evaluating the acceptance and adoption of \system{}, we used the Technology Acceptance Model (TAM) scale~\cite{venkatesh2008technology}. The whole procedure was conducted on the user's private screen to reduce bias.

Then, we conducted an interview, where we inquired about the overall usability, their feedback on the iteration process with the agent, the quality of the generated plans, and the potential future enhancements. The overall procedure took approximately 1 hour for each participant.

\ipstart{Intrinsic evaluation}
The research team evaluated how well each participant’s plan followed global exercise guidelines after both the initial planning and iteration phases. Two researchers manually read through each plan, discussed, and reached a consensus on whether each plan met the requirements of the global guidelines, as detailed in \autoref{sec:exercise_guideline}, including (1) amount (\ie, whether the total exercise time of the weekly plan exceeds 150 minutes, while counting vigorous activity double), (2) balance (\ie, whether both cardio and strength training are included in the plan), and (3) resting (\ie, whether one or more rest day(s) between exercise days are included).

\ipstart{Expert evaluation} 
To assess the generated plans from the perspective of experts, we recruited three expert planners (E1 -- E3; one male and two females) from an in-house clinic of the corporation. The experts were nationally licensed physical therapists aged between 28 and 39 ($M=31.3$), and had an average of 7 years in professional exercise planning ($SD=4.6$). We asked the experts to evaluate plans from the initial exercise planning phase both quantitatively and qualitatively, where each expert was randomly assigned six plans and asked to evaluate them.

Specifically, the experts holistically reviewed the plans as well as the constraints and conversation history, with private information masked. For each plan, they filled out our evaluation form that consists of a 7-point Likert scale of four items from the FITT principles~\cite{desimone2019tortoise}---\textit{frequency} (\ie{}, how often the exercises in the plan are), \textit{intensity} (\ie{}, how intense the exercises consisting of the plan are), \textit{time} (\ie{}, duration of the exercises consisting of the plan), and \textit{type} (\ie{}, composition of the types of exercise consisting of the plan)---a recognized and empirically validated framework consisting of salient factors in exercise plan design and assessment (1: highly unsatisfactory, 7: highly satisfactory). For each item, we also included an open-ended field asking for the rationales for the assessment.

\subsection{Analysis}

Similar to what we did for our formative study, we used a thematic analysis to code (1) participants' responses and (2) qualitative responses from expert evaluations. The two authors of this work first read and gained a sense of the raw responses independently, and each author identified emerging themes from the responses. Then, they teamed up to discuss and compare the themes during the regular meetings until they reached a consensus.


Additionally, we analyzed the interaction logs to understand the interaction between the participants and the CA. The two authors first individually reviewed the logs, linking each user action to the user-defined constraints (\ie{}, goal, availability, obstacle) and exercise type. Based on this review, we initially classified each action as \texttt{add}, \texttt{edit}, or \texttt{remove}, denoting whether it aimed to introduce a new entity, modify an existing one, or delete one. The research team then met three times to conduct a bottom-up thematic analysis to discuss and consolidate the emerging categories that could be characterized as distinct actions. Through this process, we additionally identified and defined new action types such as \textit{amount} (\ie{}, adjusting the exercise amount), \textit{question} (\ie{}, asking the agent questions), and \textit{querying exercise list} (\ie{}, requesting exercise recommendations based on user-specified constraints). These actions were then organized in a sequence for each participant.
\section{Results}\label{sec:user_study_results}

In this section, we report the results of our study in three parts: (1) collected constraints and interaction patterns with the conversational agent, (2) user evaluation of the agent and its crafted plans, and (3) quality of the generated plans.

\subsection{Collected Constraints and Interaction Patterns}

\def\goalitemwidth{0.26\textwidth}
\def\avitemwidth{0.25\textwidth}
\def\challengeitemwidth{0.36\textwidth}

\begin{table*}[]
\sffamily
	\def\arraystretch{1.15}\setlength{\tabcolsep}{0.2em}
		    \centering
\caption{Exercise goals, availabilities, and potential obstacles that \system{} surfaced from the participants during the initial planning phase}
\small
\begin{tabular}{|l|>{\raggedright\let\newline\\\arraybackslash\hspace{0pt}}p{0.28\textwidth}!{\color{lightgray}\vrule}>{\raggedright\let\newline\\\arraybackslash\hspace{0pt}}p{0.27\textwidth}!{\color{lightgray}\vrule}>{\raggedright\let\newline\\\arraybackslash\hspace{0pt}}p{0.38\textwidth}|}
\hline
\rowcolor{tableheader}\textbf{ID} & \textbf{Goal} & \textbf{Availability} & \textbf{Potential obstacle} \\
\hline
\textbf{P1} & \tablistitem{\goalitemwidth}{Weight loss}\newline\tablistitem{\goalitemwidth}{Recover energy} & \tablistitem{\avitemwidth}{Weekdays at night after 6 pm}\newline\tablistitem{\avitemwidth}{Weekends in the morning} & \tablistitem{\challengeitemwidth}{Do not wanna do exercises that heavily affect knees}\newline \tablistitem{\challengeitemwidth}{Company dinner or other appointments} \\
\arrayrulecolor{tablegrayline}\hline
\textbf{P2} & \tablistitem{\goalitemwidth}{Maintain muscular strength}\newline \tablistitem{\goalitemwidth}{Be more energetic in daily life}\newline \tablistitem{\goalitemwidth}{Weight loss}\newline \tablistitem{\goalitemwidth}{Maintain daily health}\newline \tablistitem{\goalitemwidth}{Cardio}  & \tablistitem{\avitemwidth}{After waking up}\newline \tablistitem{\avitemwidth}{If it fails, exercise afternoon or at night instead}\vspace{1mm} \newline \tablistitem{\avitemwidth}{Light exercise after lunch} & \tablistitem{\challengeitemwidth}{Light exercise at night} \newline \tablistitem{\challengeitemwidth}{Hard to exercise on the day after drinking}\newline \tablistitem{\challengeitemwidth}{Sudden schedules afternoon}\newline \tablistitem{\challengeitemwidth}{Sudden schedules at night} \\
\hline
\textbf{P3} & \tablistitem{\goalitemwidth}{Recover basic energy} & \tablistitem{\avitemwidth}{After school} & \tablistitem{\challengeitemwidth}{Difficult to exercise after heavy drinking} \\
\hline
\textbf{P4} & \tablistitem{\goalitemwidth}{Weight loss}\newline \tablistitem{\goalitemwidth}{Overcome exercise shortage since pandemic}\vspace{1mm} & \tablistitem{\avitemwidth}{Thu--Sun after 7 pm} & \tablistitem{\challengeitemwidth}{Don't want to exercise on rainy days} \\
\hline
\textbf{P5} & \tablistitem{\goalitemwidth}{Improve muscular strength}\newline \tablistitem{\goalitemwidth}{Fix posture}  & \tablistitem{\avitemwidth}{Everyday in the morning} & \tablistitem{\challengeitemwidth}{Want to exercise without equipment}\newline \tablistitem{\challengeitemwidth}{Not familiar with exercise} \\
\hline
\textbf{P6} & \tablistitem{\goalitemwidth}{Weight loss} \newline \tablistitem{\goalitemwidth}{Improve shoulder muscles} \newline \tablistitem{\goalitemwidth}{Relieve wrist pain} & \tablistitem{\avitemwidth}{Everyday in the morning except for late night}  & \tablistitem{\challengeitemwidth}{Diagnosed with right shoulder subluxation} \\
\hline
\textbf{P7} &\tablistitem{\goalitemwidth}{Recover energy}\newline \tablistitem{\goalitemwidth}{Weight loss}\newline \tablistitem{\goalitemwidth}{Improve muscles} & \tablistitem{\avitemwidth}{Weekdays in the morning \& at night} & \tablistitem{\challengeitemwidth}{Kids' day off from school or appointment}\newline \tablistitem{\challengeitemwidth}{Kids/husband come back home early} \\
\hline
\textbf{P8} & \tablistitem{\goalitemwidth}{Weight loss} \newline \tablistitem{\goalitemwidth}{Recover energy}\newline \tablistitem{\goalitemwidth}{Relieve stress} \newline \tablistitem{\goalitemwidth}{Get hobbies} & \tablistitem{\avitemwidth}{Weekdays in the morning}\newline \tablistitem{\avitemwidth}{Weekdays afternoon} \newline \tablistitem{\avitemwidth}{Weekends at any time} & \tablistitem{\challengeitemwidth}{Difficult to exercise after drinking or sleeping late} \newline \tablistitem{\challengeitemwidth}{Postpone exercise if there is a schedule with others} \\
\hline
\textbf{P9} & \tablistitem{\goalitemwidth}{Improve swimming skills} \newline \tablistitem{\goalitemwidth}{Improve muscular strength} & \tablistitem{\avitemwidth}{Weekdays in the morning} \newline \tablistitem{\avitemwidth}{Unable to exercise on Mon--Fri as already doing swimming}\vspace{1mm} & \tablistitem{\challengeitemwidth}{Difficult to exercise if a kid is sick} \\
\hline
\textbf{P10}  & \tablistitem{\goalitemwidth}{Weight loss} \newline \tablistitem{\goalitemwidth}{Recover energy} & \tablistitem{\avitemwidth}{Weekdays after school at night} \newline \tablistitem{\avitemwidth}{Weekends afternoon} \newline \tablistitem{\avitemwidth}{Tuesday afternoon--night} & \tablistitem{\challengeitemwidth}{Sleepy after school} \\
\hline
\textbf{P11}  & \tablistitem{\goalitemwidth}{Weight loss} \newline \tablistitem{\goalitemwidth}{Cardio} & \tablistitem{\avitemwidth}{After dinner} & \tablistitem{\challengeitemwidth}{Location constraint} \\
\hline
\textbf{P12}  & \tablistitem{\goalitemwidth}{Weight loss} & \tablistitem{\avitemwidth}{Everyday in the morning \& at night} & \tablistitem{\challengeitemwidth}{Diagnosed with back disc} \\
\hline
\textbf{P13}  & \tablistitem{\goalitemwidth}{Weight loss} & \tablistitem{\avitemwidth}{Three times per week in the morning (9--12 am)}  & \tablistitem{\challengeitemwidth}{Prefer indoor exercise} \newline \tablistitem{\challengeitemwidth}{Diagnosed with peripheral edema}  \\
\hline
\textbf{P14}  & \tablistitem{\goalitemwidth}{Weight increase} \newline \tablistitem{\goalitemwidth}{Recover energy} & \tablistitem{\avitemwidth}{Everyday after 7 pm except for Sat} & \tablistitem{\challengeitemwidth}{Want to avoid excessively using the right index finger} \\
\hline
\textbf{P15}  & \tablistitem{\goalitemwidth}{Improve arm muscles} \newline \tablistitem{\goalitemwidth}{Want to make waist look thinner}\vspace{1mm} & \tablistitem{\avitemwidth}{Weekdays at night} \newline \tablistitem{\avitemwidth}{Weekends 10--12 am} & \tablistitem{\challengeitemwidth}{Weekday night party}\newline \tablistitem{\challengeitemwidth}{Wish to exercise three times per week} \\
\hline
\textbf{P16}  & \tablistitem{\goalitemwidth}{Weight loss} \newline \tablistitem{\goalitemwidth}{Relieve waist pain} \newline \tablistitem{\goalitemwidth}{Get broad shoulders} & \tablistitem{\avitemwidth}{Tue--Thu after school} \newline \tablistitem{\avitemwidth}{Fri \& Sat before work} \newline \tablistitem{\challengeitemwidth}{Sun \& Mon at anytime} & N/A (provided no obstacle)\\
\hline
\textbf{P17}  & \tablistitem{\goalitemwidth}{Improve golf--backswing skills} & \tablistitem{\avitemwidth}{Mon at anytime}\newline \tablistitem{\avitemwidth}{Thu \& Fri at night} & \tablistitem{\challengeitemwidth}{Economical exercises} \\
\hline
\textbf{P18}  & \tablistitem{\goalitemwidth}{Recover energy} \newline \tablistitem{\goalitemwidth}{Improve muscles} \newline \tablistitem{\goalitemwidth}{Relieve back pain} & \tablistitem{\avitemwidth}{After work}\newline \tablistitem{\avitemwidth}{Weekends afternoon} & N/A (provided no obstacle)\\
\arrayrulecolor{black}\hline
\end{tabular}
\Description{This table outlines the constraints identified by participants in our study through their interactions with our conversational agent. Each row includes an ID, goal, availability, and potential obstacles, with various constraints for participants P1 to P18 listed.}
\label{tab:constraints}
\end{table*}

As shown in \autoref{tab:constraints}, through interacting with the agent, participants provided a wide range of constraints related to their lifestyle for crafting a personalized plan. On average, participants shared 2.28 exercise goals ($SD=1.04$), 1.72 availabilities ($SD=0.80$), and anticipated 1.33 potential obstacles ($SD=0.88$). Some common goals that the participants described include weight loss ($N=11$), recovering daily energy ($N=8$), and maintaining/improving muscular strength ($N=5$). For availability, only 5 participants described their availability in the exact time format (\eg{}, after 7 pm); the others described all of their availabilities freely in a descriptive form (\eg{}, after school). Lastly, participants described their potential obstacles in a highly personalized expression by drawing connections to various aspects of their own lifestyles and circumstances, such as heavy drinking (P2, P3), kid's schedule (P7, P9), and party~(P15).

\begin{figure*}
    \centering
    \includegraphics[width=\textwidth]{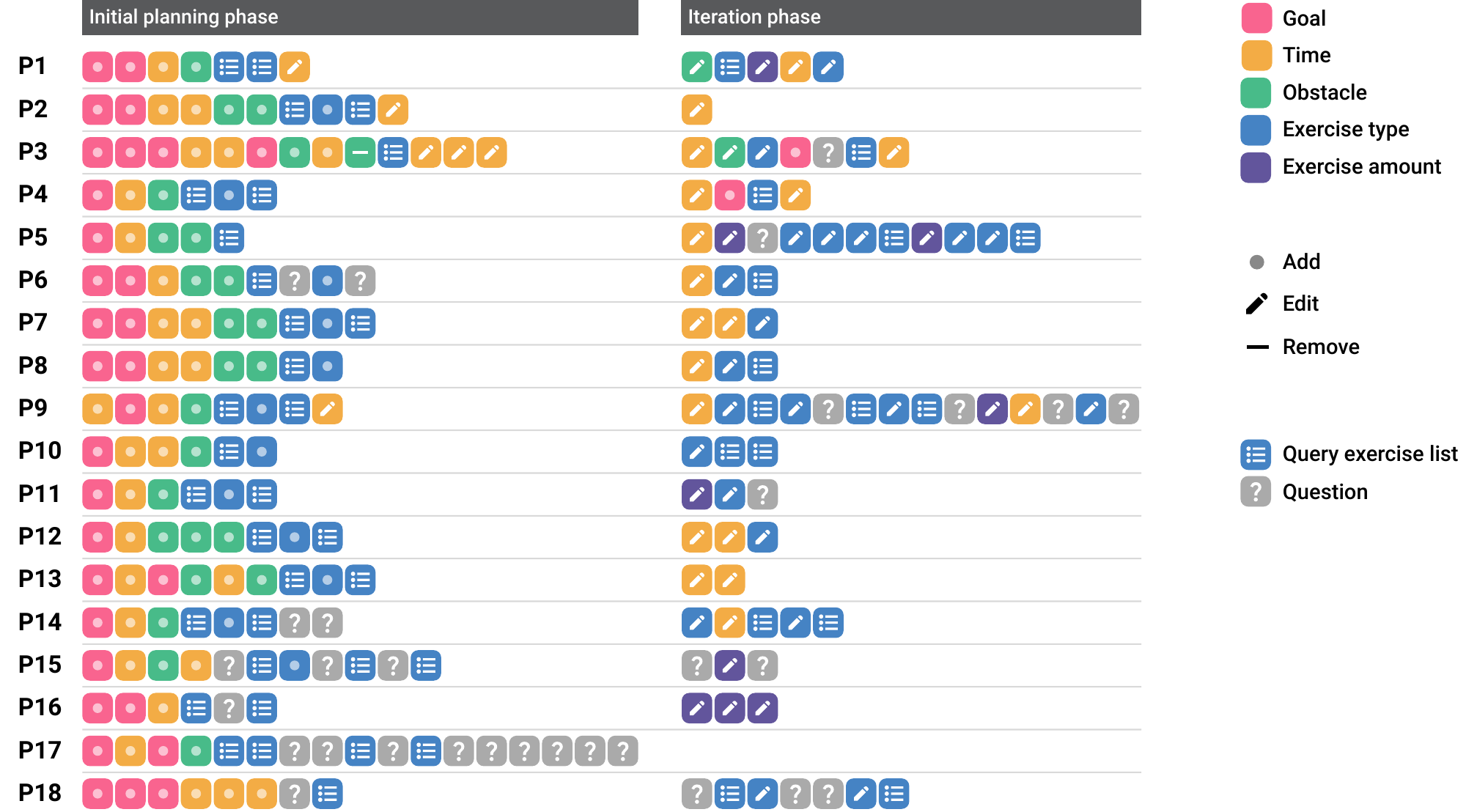}
    \caption{Sequence of how the participants interacted with the conversational agent to tailor their exercise plan}
    \Description{This figure depicts the sequence of edits made by participants as they interacted with the conversational agent to customize their exercise plans during our exploratory study. The edits include categories such as goal, time, obstacle, exercise type, and exercise amount, with types of edits classified as add, edit, or remove. Additionally, participants queried exercise lists and posed questions as additional interactions.}
    \label{fig:sequence}
\end{figure*}

\autoref{fig:sequence} illustrates how these constraints were provided to the conversational agent and modified for each participant across the two study phases. In the early stage, participants generally followed the sequence of information that the conversational agent's dialogue analyzer was instructed to collect. As the interaction progressed, they iterated on their constraints in individual ways through flexible conversational interaction. Ten participants (56\%) also asked questions (\autoref{fig:sequence}; gray rectangles) to \system{} about exercise and other relevant topics.

\subsection{User Evaluation}

\autoref{tab:user-eval-summary:tam} illustrates the technology acceptance scales, and \autoref{tab:user-eval-summary:personalization} and \ref{tab:user-eval-summary:actionability} illustrate the distribution of user evaluation assessing the quality of plans guided by \system{} (\ie{}, personalization, actionability), respectively. Below, we describe the participants' quantitative evaluation in detail, along with their qualitative feedback, offering insights into the role of the agent in creating quality plans and the user perception of CA-supported exercise planning.

\subsubsection{Perceptions of the use of conversational agent}

Participants indicated a positive inclination towards interacting with the agent, where the perceived usefulness received a rating of 5.43 on average ($SD=0.99$). Similarly, participants found \system{} to be easy to use, with an average of 6.00 ($SD=1.12$). As for the intention to continue using it, participants responded with an average rating of 5.52 ($SD=1.26$).

From the interview, we could uncover factors that contributed to the participants' intention to keep using it. First, free-flowing and flexible conversations enabled by the LLM-driven conversational agent allowed participants to freely initiate actions (\eg{}, introducing additional constraints) whenever they naturally came to mind, not only limited to when prompted by the agent's questions. This flexibility enhanced their perception of the agent's utility in the planning process: \textit{“Even if I suddenly went back to a previous question or said something else, the system seamlessly continued the conversation which made the chatting more convenient.”}~(P3)

Second, displaying the conversational history in a dashboard alongside the chat panel allowed participants to easily track the constraints they had specified. This visibility helped them adjust plans efficiently without needing to manually navigate through the entire chat history. With a clear overview, participants could stay aware of all constraints and correspondingly updated plans without having to read through lengthy dialogues, increasing their confidence in using the agent continuously: \textit{“The dashboard neatly organizes and updates the information every time I entered constraints, which I find very convenient. Often times when I plan things like this, I have to make separate notes on my phone, right? Now I can just provide it to the agent, and it automatically organizes it for me (...) I consider this as a very useful component.”}~(P12)

At the same time, participants also anticipated the future integration of additional contextual information that the agent could collect and consider during the personalization process. For instance, some participants suggested that incorporating context-aware features, such as providing exercise recommendations based on their current location and weather conditions, could significantly enhance the system's utility. Additionally, soliciting more detailed constraints from participants, such as whether they have children~(P13) or specific muscle areas requiring rehabilitation~(P15), was identified as a future enhancement that would further enhance the usefulness of the agent.

\subsubsection{Personalization}
Overall, participants found that interacting with \system{} helped them to generate personalized plans. They reported the plans to fit their personal lifestyle ($M=6.00$, $SD=0.77$), and that they were generally likely to follow the plans ($M=5.83$, $SD=0.99$), as evaluated on a 7-point Likert scale (1: strongly disagree, 7: strongly agree).

Participants reported that the expressivity of the conversational agent when guiding them, as well as its understandability of user requests in real-time, greatly contributed to successfully tailoring exercise plans according to their preferences and constraints. In contrast to constrained expert-client communication settings, \system{} allowed participants to make unlimited requests easily through natural language. This served as the flexibility to personalize their plans as extensively as they desired: ``\textit{It was refreshing to have schedules tailored to my personal time and listen to my request. I was really surprised to see that it could do that well.}''~(P15) Observing the output plans where the agent successfully incorporated these personal requests and constraints in real-time, participants found the planning process with \system{} to be highly personalized: ``\textit{Once I requested, it extended the duration of each exercise session by 15 minutes.}''~(P15); ``\textit{I said I wanted to do simple, sweat-free, and noiseless exercises at home. Tailoring my plan using this, it was great to see that my preferences and conditions were reflected exactly in the plan that looks easy to follow.}''~(P10)

Not limited to the initial generation of plans, participants also highlighted the agent's potential to let users reiterate their existing plans and constraints over time. For instance, when unforeseen changes occur, which could make it difficult to adhere to the original plans, \system{}'s capability of facilitating iterative planning would greatly help them quickly adapt to unexpected changes: ``\textit{It is really nice having the option to easily modify the existing exercise plan when a new goal arises (...) For example, if I suddenly injure my leg and need rehabilitation, I'm sure it would also be well-reflected in my plan.}''~(P16); \textit{``If there's a change in my availability, being able to make adjustments instantly like this, I believe I would use it frequently.''}~(P14) With such support that enables users to freely iterate on their plans, participants came up with various potential use cases of \system{}, such as finding and engaging in lightweight exercises that can be done on the go or when they suddenly have some free time: \textit{“Let's assume that I want to utilize some spare moments, for example, when I finish lunch early and have about 20-30 minutes left. Then I could easily use this system in my workplace to use those spare moments.”}~(P5)

\begin{table*}[t!]
\sffamily
\def\arraystretch{1.2}\setlength{\tabcolsep}{0.35em}
		    \centering
    \caption{Mean participant ratings with standard deviation (7-point scale) across evaluation categories.}
    \begin{tabular}{|c!{\color{lightgray}\vrule}c!{\color{lightgray}\vrule}c|c!{\color{lightgray}\vrule}c|c!{\color{lightgray}\vrule}c!{\color{lightgray}\vrule}c!{\color{lightgray}\vrule}c|}
    \hline
    \rowcolor{tableheader}
    \multicolumn{3}{|c|}{\textbf{(a) Technology acceptance}} & 
    \multicolumn{2}{c|}{\textbf{(b) Personalization}} & 
    \multicolumn{4}{c|}{\textbf{(c) Actionability}} \\
    \hline
    \textit{\textbf{Usefulness}} & \textit{\textbf{Ease of use}} & \textit{\textbf{Intention to use}}
    & \textit{\textbf{Follow}} & \textit{\textbf{Fit}} 
    & \textit{\textbf{Specificity}} & \textit{\textbf{Encouragement}} & \textit{\textbf{Vocabulary}} & \textit{\textbf{Accuracy}} \\
    \hline
    \(5.43 \pm 0.99\) & \(6.00 \pm 1.12\) & \(5.52 \pm 1.26\) & 
    \(5.83 \pm 0.99\) & \(6.00 \pm 0.77\) & 
    \(5.06 \pm 1.51\) & \(5.56 \pm 1.34\) & \(6.19 \pm 1.11\) & \(5.72 \pm 1.23\) \\
    \hline
    \end{tabular}
    \label{tab:user-eval-summary}
    \labelphantom{tab:user-eval-summary:tam}
    \labelphantom{tab:user-eval-summary:personalization}
    \labelphantom{tab:user-eval-summary:actionability}
    \Description{Average participant ratings for each sub-questionnaire under technology acceptance, personalization, and actionability.}
\end{table*}

\subsubsection{Actionability}

As in \autoref{tab:user-eval-summary:actionability}, participants were also positive about the actionability of the plans they created with \system{}, which were generally received to be specific with enough details to act upon~($M=5.06$, $SD=1.51$). Participants also found the agent's presentation of the plan and its accompanying information encouraging~($M=5.56$, $SD=1.34$), described with straightforward vocabulary~($M=6.19$, $SD=1.11$), and accurate~($M=5.72$, $SD=1.23$), responded on a 7-point Likert scale (1: strongly disagree, 7: strongly agree).

Our qualitative analysis revealed that participants found the \texttt{IF-THEN} implementation intention format presented by the agent practical and adaptable, especially for individuals with fluctuating schedules. Avoiding vague timing instructions (\eg{}, 3 times per week) or rigid time constraints (\eg{}, 7 pm) while contextualizing the plan to the user-described situations, the plan was perceived as realistic and easy to remember, making participants find it to be more actionable: ``\textit{I think it's better when it tells you to do some exercise based on the situation like this. Honestly, sticking to a set time isn't always easy to follow through with, in reality.}''~(P6)

Furthermore, participants reported the plans guided by the agent to be well-adhering to the specific constraints they provided: \textit{“For every information I added to the chat, the system successfully reflected those to my exercise plans.”}~(P3) The plans were also reported to be presented in sufficient detail to follow by specifying the exercise type and amount, which was viewed as clear and easy to follow: \textit{``What surprised me was how it instructed me on what to do on each day, like there was a clear outline. I liked that it was so specific. I tend to prefer clear instructions (...) Nowadays, there are just too many choices, and I tend to dislike making decisions. So, having such clear instructions made me appreciate why I should use this and why I rated it highly.''}~(P4) With such specificity of the plans, participants noted that the generated plans are systematic and presented in an actionable format: \textit{``I felt like I could systematically handle various types of exercises a bit better. It gave me a feeling of being well-grounded.''}~(P6)

Participants also highlighted that providing coping plans for each exercise plan contributed further to its actionability. They expected that, even when facing obstacles that might lead them to skip an exercise, these coping plans would serve as clear guidance to make up for the exercise: \textit{``If I find myself unable to do my exercise and I'm debating whether to skip it for the day, seeing this alternative [coping plan] might make me think, `Well, if I can't follow the original plan, I might as well do the alternative one today,' and it would induce to start exercising anyway.''}~(P18)

\subsection{Evaluation of the Quality of Generated Plans}

\subsubsection{Intrinsic evaluation}\label{sec:adhere}

From our intrinsic evaluation, we found that \system{} effectively guided participants in aligning their plans with the global guidelines, as outlined in \autoref{sec:global_guidelines}:

\ipstart{(i) Amount}
\system{} successfully met the amount guideline while calibrating amounts across individual sessions. Of the initial plans, 15 of 18 (83\%) reached the required weekly total; two participants received less as they mentioned they were already personally engaging in other activities outside the planning, and one participant manually asked to exclude a session. In the iteration phase, all but three participants (including two from the initial phase) finalized plans that satisfied the amount guideline.

\ipstart{(ii) Balance}
\system{} allows users to first choose their preferred exercise types, while suggesting adding complementary types if the exercise types from only one category (cardio or strength) were chosen. In the initial planning phase, nine participants selected either cardio or strength only, prompting \system{} to recommend adding the missing type, where eight of the nine accepted to create a balanced routine. During the iteration phase, one participant provided an injury-related constraint, leading \system{} to remove certain exercises and omit one type.

\ipstart{(iii) Resting}
Of all, 14 plans (78\%) in the initial phase met the guideline for ensuring a rest day. In the exceptional four cases, the system could not include this gap as participants manually indicated the days of the week for exercise. In the iteration phase, three more participants requested schedule changes based on their scenario which necessitated consecutive exercises; for all the others who did not add such inevitable constraints, the iterated plan satisfied the guideline.

\subsubsection{Expert evaluation}

Expert planners generally found the plans generated by \system{} to be well-adhering the FITT principle---how adequately the \textit{frequency}, \textit{intensity}, \textit{time}, and \textit{type} of exercises were formulated (see \autoref{tab:expert-fitt-summary}). Below, we describe the assessment and feedback we gained from the experts and the potential room for further enhancing the plans.

\ipstart{(i) Frequency} Experts generally rated the exercise frequency of the plans as well-defined, averaging 5.67 on a 7-point Likert scale ($SD=1.53$). Qualitative feedback highlighted the agent's success in accommodating both the 150 minutes per week guideline and individual preferences, particularly its approach of evenly distributing exercise throughout the week while incorporating rest days: ``\textit{It's highly commendable to reflect the exercise guideline by scheduling exercise with the assigned time for at least 3 times a week and incorporating the concept of rest on the day after exercise.}''~(E1)  However, experts also suggested future improvements, including prompting the agent to adjust frequency based on the number of exercise types selected, potentially increasing frequency for plans corresponding to the number of exercises the participants wish to do (E3:\textit{“Given the four different exercise types (that the participant mentioned they wished to do), it may make sense to increase the exercise frequency from the current four times a week to five or six.”}) and decreasing frequency for plans with similar exercises to avoid muscle fatigue and injury risk (E2: \textit{“The plan consists of 7 days of exercise sessions that target the abdomen and lower body, which could potentially lead to muscle fatigue. It's essential to reduce the frequency.”})

\begin{table}[b]
\sffamily
	\def\arraystretch{1.15}\setlength{\tabcolsep}{0.5em}
		    \centering
    \caption{Mean expert ratings with standard deviation (7-point scale) across FITT principles.}
    \begin{tabular}{|c!{\color{lightgray}\vrule}c!{\color{lightgray}\vrule}c!{\color{lightgray}\vrule}c|}
    \hline
    \rowcolor{tableheader}
    \textbf{Frequency} & \textbf{Intensity} & \textbf{Time} & \textbf{Type} \\
    \hline
    \(5.67 \pm 1.53\) & \(4.28 \pm 1.32\) & \(5.06 \pm 1.80\) & \(3.89 \pm 1.45\) \\
    \hline
    \end{tabular}
    \label{tab:expert-fitt-summary}
    \Description{Average expert planner ratings for each FITT component: frequency, intensity, time, and type of exercises generated by the agent.}
\end{table}

\ipstart{(ii) Intensity} The experts rated the exercise plan's intensity with a general favorability with room for improvement, with an average score of 4.28 ($SD=1.32$). They praised the system for preventing intensity-related issues through coping plans based on participant-reported obstacles, such as advising participants with back pain to stop exercising and consult specialists if needed: \textit{“I found cautionary comments for the patients with back pain to be great, along with the appropriate intensity of exercise offered.”}~(E2). However, experts suggested enhancements to the system's guidance on intensity. The agent currently recommends increasing the amount of exercise as a progression measure if users are satisfied with previous plans. On top of the time, E1 suggested that the intensity of the plans can also be used as a measure for the progression: \textit{“In terms of the intensity of this plan, I consider it appropriate. Given that the participant is healthy, I also recommend the user start with moderate intensity and gradually progress to higher intensity.”} Similarly, the agent uses predefined intensity information in our predefined exercise list to guide recommendations, but based on the user needs like weight loss or muscle strength improvement, E1 suggested customization to include high-intensity exercises corresponding to participants' individual goals: \textit{“To achieve weight loss, I believe it is necessary to include high-intensity aerobic exercises that have a higher level of intensity.”}

\ipstart{(iii) Time} As detailed in \autoref{sec:adhere}, the agent effectively generated exercise plans that comply with the ACSM guidelines for exercise time. The evaluation of these plans particularly praised the adherence to the time component, with a rating of 5.06 ($SD=1.80$). Experts expressed satisfaction with the planning, and offered recommendations to improve flexibility. For example, if a user is unable to commit to a 30-minute exercise, E1 suggested it could be further broken into shorter sessions (\eg{}, three 10-minute sessions) for flexible planning: \textit{“I think the amount of time has been planned well. If the client is unable to commit to a 30-minute exercise, you can also advise them to break it down into three 10-minute sessions.”}~(E1) Additionally, there is potential to enhance \system{} by operationalizing exercise time not just in weekly totals but also in per-session durations. While the system meets the ACSM guidelines for total weekly duration, experts identified areas for improvement in individual sessions. For example, if a user has limited time for exercise, the agent currently generates long, higher-intensity sessions to meet the guidelines within fewer available days. However, planners cautioned that such prolonged, intense sessions could lead to overexertion, advising against these exceptional cases: \textit{“For the case of high-intensity exercises, prescribing a 50-minute session of strength training is excessive for the participants.”}~(E2)

\ipstart{(iv) Type} The exercise types within the plans received a slightly below satisfactory rating of 3.89 ($SD=1.45$), emphasizing the need for enhanced tailoring. Expert feedback highlighted key areas for improvement, particularly in guiding users to balance cardio and strength exercises; while the agent already encouraged users to include at least one exercise of the opposing type, it did not enforce equal distribution, resulting in some plans being heavily skewed or sometimes omitting one category. To address this, E2 suggested adopting a more assertive tone when presenting recommendations to ensure balanced planning: \textit{“Only the exercises the user wanted to do were included. However, as this is an interaction where AI sets exercise goals together with the participant, ‘necessary exercises’ should also be guided.”} Also, experts identified inaccuracies when specific muscle groups were not explicitly mentioned in participants’ goals—such as the agent relying on cosine similarity between “golf” and exercise descriptions, which led to overlooking beneficial strength and flexibility routines. E3 pointed out this limitation, suggesting that enhancing \system{} to infer relevant muscle groups, even when not explicitly stated, could improve the accuracy of exercise recommendations: \textit{“Other exercises that could enhance golf performance were not adequately suggested (...) recommendations for improving golf backswings should include exercises that enhance flexibility, core strength, and lower body strength.”}
\section{Discussion}
In this section, we discuss lessons learned from designing and implementing a CA for personalized exercise planning, as well as its evaluation from our user study.

\subsection{Leveraging LLM-driven CAs for Exercise Planning}

Our work proposed leveraging LLM-driven CAs to create personalized exercise plans that account for individual constraints, while aligning the plans with global guidelines. Instead of relying solely on simple LLM generation based on the knowledge base of generic models---which may be prone to hallucination~\cite{huang2023survey} and lack of the output's alignment with the real-world practice and guidelines~\cite{liu2023trustworthy}, we developed a pipeline that integrates expert-verified exercise lists and guidelines to inform the generated plans in a way better aligned with real-world practices, which received positive feedback in expert evaluations. Additionally, visualizing the current planning status on a dashboard helped participants better keep track of their plans, without losing the context while engaging in the back-to-back conversation.

Building on these, the free-form conversations carried by LLM-driven CAs enabled participants to provide their exercise-related constraints intuitively and flexibly, resulting in the system identifying diverse and unique constraints from participants (\cf,~\autoref{tab:constraints}). Additionally, the conversational interaction allowed the exchange of questions and reiterations of the plans (\cf,~\autoref{fig:interaction}), seamlessly interleaved in the user interface. This was shown to be effective during the iteration phase of the study, where \system{} successfully adjusted the plans per user requests. Observing the system reflect their requested edits in the plan, participants expressed intention to use \system{} in the long term and frequently throughout their exercise journey. Our work suggests that LLM-driven conversational interaction could successfully simulate natural interactions in exercise planning settings, while demonstrating opportunities for long-term engagement with an exercise assistant agent.

In this process, unlike typical open-domain conversations where most LLM-driven CAs operate, \system{} needed to reliably adhere to user-defined constraints and create exercise plans grounded in established guidelines. To achieve this, we employed several design choices to enhance compliance with our design goals. First, we incorporated two distinct agent routines: one dedicated to generating conversational dialogue and another for transforming user dialogue into formatted data, used for input summaries and exercise plan generation. Having two routines dedicated to conversation and analysis respectively, our CA could attain reliability in both tasks. Second, to enable the agent to reference external exercise knowledge, we implemented a retrieval-augmented generation technique to integrate an existing exercise database, allowing for more evidence-based planning while preserving the flexibility of agent-driven conversations. Additionally, this approach would allow \system{} to easily tailor its focus on specific exercise environments (\eg{}, bodybuilding) or organizational settings where clients have access to only a restricted set of exercises, simply by replacing or modifying the exercise database.

\subsection{Incorporating Nuanced Perspectives of Domain Experts}

Although \system{} generally complied with the exercise guidelines as intended, evaluation from the expert also revealed the future enhancements for some components of the crafted plans (\ie{}, exercise intensity, types), suggesting edits that they would have applied to the plans based on their own hands-on experiences (\eg{}, recommending a certain exercise intensity for achieving specific exercise goals). This points out that, although \system{} is reported to successfully take into account various individualized factors and exercise guidelines during the exercise planning process, human expertise may still contribute to enhancing the quality and effectiveness of the plans. Since such edits and potential contributions may be grounded upon the experts' tacit knowledge from the lessons they learned over time, it is not trivial to formalize such knowledge into global guidelines and reflect them to the agent's instruction.

A promising way to address this gap is through multi-agent collaboration, where multiple conversational agents (CAs) embody distinct expert personas. For example, our study surfaced a key tension between maximizing exercise performance and preventing overexertion. To navigate such trade-offs, future systems could employ multiple agents (\eg{}, a progressive planner focused on performance gains vs. a preventive planner emphasizing injury avoidance) that critique and refine plans from their respective perspectives. This setup would help users explore alternative viewpoints and make more informed decisions, addressing nuances that a single-agent model might overlook. This deliberative, agentic workflow has also been attempted in various domains to integrate complementary expertise and mirror real-world expert collaboration through discussion, negotiation, and consensus~\cite{chen2023agentverse, talebirad2023multiagent}. By synthesizing diverse perspectives, we hypothesize that such systems involving multiple agents with diverse viewpoints could generate more balanced plans and enhance informed decision-making.

\subsection{Generalizability to Other Planning Domains}

One key aspect of our system was the integration of implementation intentions, where the users are provided with \texttt{IF-THEN} statements linked to their availabilities collected through chatting with the CA. From the study, we identified that the participants perceived such situation-based expressions as highly comprehensible and adaptable, compared to vague amount-based or rigid time-based instructions. Similarly, as such implementation intention strategies have been shown effective in a variety of behavior change tasks (\eg{}, diet control~\cite{adriaanse2011implementation, achtziger2008implementation, gratton2007promoting, reuter2008dietary}, smoking cessation~\cite{mcwilliams2019beyond, conner2010long}), we posit that our approach is also adaptable to various other behavior change contexts. Particularly, since our system is composed of a set of easy-to-alter instructions in a natural language that define the constraints to be collected, we believe that the adaptation process for various other tasks can be significantly straightforward, requiring minimal changes to tailor these instructions to reflect the domain-specific constraints of each new context.

\subsection{Towards a Long-term Interaction with \system{}}

From our formative study, we identified that exercise planning is an iterative process that takes place in the long term as the user's exercise progresses. Motivated by this, our exploratory user study simulated such an iteration and revealed opportunities for the CA as a long-term exercise companion. To further support the interaction of users with the CA in the extended duration, future works need to longitudinally study how the system can be expanded to help elicit information from users over time, and how to leverage that information to inform future revisions of the plan.

As the system scales up, we believe that incorporating context-aware features would help the plans to be even more aligned with the user-provided constraints, assisting in generating more realistic and customized plans. For instance, integrating location-aware exercise recommendations could enable \system{} to take into account factors driven by real-time information, such as weather conditions, nearby exercise facilities, or nearby routes that allow users to perform exercise on the go (\eg{}, a specific route for running while going back home). Such a level of contextualization would make the generated plans even more closely connected to the user's real-world situation and make the exercise plans more engaging. Similarly, other features that reflect an up-to-date health status of the user could be incorporated into future revisions of \system{} to create even richer and more personalized exercise planning.
\subsection{Limitations and Future Work}


While our findings demonstrate the promise of LLM-driven conversational agents in personalizing exercise plans, several limitations remain, suggesting important directions for future work. First, our recommendations are based on a curated dataset that, while expert-vetted, may not reflect the full range of exercise types or cultural preferences. Future work should explore expanding the exercise corpus with more diverse, representative data sources, and develop mechanisms to detect and mitigate potential biases in both the dataset and model outputs.

Second, despite the use of retrieval-augmented generation and rule-based prompting, the LLM occasionally generated plans lacking nuance, such as missing rest days or offering overly general suggestions. These limitations point to the need for integrating more domain-specific reasoning or constraint satisfaction approaches alongside LLMs---via hybrid models or fine-tuning with expert-reviewed exercise prescriptions.

Third, our study focused on short-term interactions; long-term adherence, motivation, and engagement with AI-generated plans remain open questions. Future research should investigate how systems like \system{} perform in real-world, longitudinal deployments with usage over weeks or months, and explore interventions to sustain user motivation (\eg{}, adaptive check-ins, habit formation scaffolds, social accountability features).

Finally, ethical and privacy considerations need to be further explored. Users may over-trust or misinterpret AI recommendations, especially in sensitive domains like health. Future versions should incorporate transparency mechanisms (\eg{}, rationale generation, uncertainty estimation), offer opt-in controls over data sharing, and support human-in-the-loop oversight, ensuring that the agent is positioned as a supportive assistant rather than an authority.
\section{Conclusion}

In this study, we propose \system{}, an LLM-driven conversational agent that helps users create personalized exercise plans through natural dialogue. Based on a user study ($N=18$) and evaluation of the generated plans, we highlighted \system{}'s potential to guide personalized, guideline-informed exercise planning. We also discuss design implications for improving LLM-driven conversational agents in personalized exercise planning.

\begin{acks}
We would like to thank Elena Agapie for generously providing the exercise dataset used in \system{}'s exercise retrieval. We also thank participants from our formative interviews and user study for their time and effort. This work was supported through a research internship at NAVER AI Lab of NAVER Cloud.
\end{acks} 

\bibliographystyle{ACM-Reference-Format}
\bibliography{bibliography}

\appendix
\onecolumn
\section{LLM Instructions} \label{apdx:instructions}

\subsection{Base Prompt for Dialogue Analyzer} \label{apdx:dialogue_analyzer}

\small\texttt{- Analyze the input dialogue and return an array of JSON objects each of which denotes an update for this summary object.\\- The user may mention multiple entities, such as goals and obstacles, or corrections to previous entities.\\- You are allowed to use the following set of methods for update:\\\{\\ \hspace*{3mm}target: "goal" | "availability" | "obstacle" | "recommended\_exercise" | "implementation\_intention",\\ \hspace*{3mm}method: "add" | "update" | "remove"\\\hspace*{3mm}params: \{ // for update\\\hspace*{6mm}id: string,\\\hspace*{6mm}update: \{\} // will be overwritten to the corresponding element.\\\hspace*{3mm}\} | \{ // for addition\\\hspace*{6mm}entity: \{\} // a new entity without ID; ID will be assigned by the system. Only for implementation\_intention, assign a random ID in case you use the "parent\_ids" property.\\\hspace*{3mm}\} | \{ // for removal\\\hspace*{6mm}id: string\\\hspace*{3mm}\}\\\}\\---\\ If there is nothing to be updated, return [].}

\normalsize
\section{Study Details}

\begin{table}[h]
\centering
\caption{Demographic information of participants in the formative study.}
\small
\begin{tabular}{llll}
\toprule
\textbf{PID} & \textbf{Age} & \textbf{Gender} & \textbf{Occupation} \\
\midrule
FC1 & 44 & Female & Homemaker \\
FC2 & 26 & Female & Full-time employee (software engineer) \\
FC3 & 30 & Female & Freelancer \\
FC4 & 56 & Male   & Retired \\
FC5 & 29 & Female & Full-time employee (software engineer) \\
FC6 & 46 & Female & Homemaker \\
FC7 & 31 & Male   & Graduate student \\
FC8 & 32 & Female & Full-time employee (product manager) \\
\bottomrule
\end{tabular}
\end{table}

\begin{table}[h]
\centering
\caption{Demographic information of participants in the main user study.}
\small
\begin{tabular}{lllll}
\toprule
\textbf{PID} & \textbf{Age} & \textbf{Gender} & \textbf{Occupation} & \textbf{\leftcell{General goal for\\exercise}} \\
\midrule
P1 & 39 & Female & Full-time employee & Diet \\
P2 & 54 & Male & Retired & \leftcell{Improve fitness,\\maintain muscle mass} \\
P3 & 19 & Male & Undergraduate student & \leftcell{Increase muscle mass,\\improve fitness} \\
P4 & 45 & Female & Homemaker & Manage blood pressure \\
P5 & 44 & Female & Homemaker & Posture correction \\
P6 & 20 & Male & Undergraduate student & Diet, improve fitness \\
P7 & 46 & Female & Homemaker & \leftcell{Improve fitness, increase\\muscle mass, diet} \\
P8 & 25 & Female & \leftcell{Part-time employee\\(customer service)} & Improve fitness, diet \\
P9 & 32 & Female & Homemaker & Improve fitness \\
P10 & 19 & Female & Undergraduate student & Improve fitness, diet \\
P11 & 27 & Female & \leftcell{Part-time employee\\(sales)} & Diet \\
P12 & 30 & Male & \leftcell{Part-time employee\\(customer service)} & Diet \\
P13 & 38 & Female & Homemaker & Diet \\
P14 & 22 & Male & Undergraduate student & Increase muscle mass \\
P15 & 24 & Female & Undergraduate student & Increase muscle mass \\
P16 & 20 & Male & Undergraduate student & Diet \\
P17 & 48 & Male & \leftcell{Full-time employee\\(public sector)} & Health management \\
P18 & 46 & Female & \leftcell{Part-time employee\\(tax)} & Increase muscle mass \\
\bottomrule
\end{tabular}
\end{table}

\end{document}